\documentclass{cjpsuppl}

\usepackage{epsfig}

\begin{document}
\title{Status of SUSY Searches}
\authori{J.-F. Grivaz}
\addressi{Laboratoire de l'Acc\'el\'erateur Lin\'eaire, Orsay, France}
\authorii{}    \addressii{}
\authoriii{}   \addressiii{}
\authoriv{}    \addressiv{}
\authorv{}     \addressv{}
\authorvi{}    \addressvi{}
\headtitle{Status of SUSY Searches}
\headauthor{J.-F. Grivaz}
\lastevenhead{J.-F. Grivaz: Status of SUSY Searches}
\pacs{\tt xx.xx}
\refnum{}
\daterec{31 October 2004;\\final version 31 October 2004}
\suppl{A}  \year{2004} \setcounter{page}{1}

\newcommand{\met}{\mbox{\ensuremath{\,\slash\kern-.7em E_{T}}}}

\maketitle
\begin{abstract}
Searches for supersymmetric particles at LEP are first reviewed,
with some emphasis on the mass limit for the lightest neutralino. 
The framework 
is the MSSM with R-parity conservation, with brief excursions towards mSUGRA 
and GMSB. Next, recent results obtained in the Run II of the Tevatron are 
presented.
\end{abstract}


\section{Introduction}

Let me first start with what this talk will not be:
\begin{itemize}
\item an introduction to supersymmetry. First of all you are all experts 
in the field, and second we just attended an excellent introductory 
presentation\cite{kraml};
\item a discussion of the cosmological aspects, mostly due to my inability;
\item a comprehensive review because of the limited amount of time available. 
\end{itemize}
This discussion will rather focus on:
\begin{itemize}
\item the LEP legacy, with minimal experimental details, because the LEP 
results often remain the most constraining at this point;
\item recent results from the Run II of the Tevatron, which is curently
the most powerful collider in operation. At the Tevatron, proton-antiproton 
collisions take place at a center-of-mass energy of 1.96~TeV, and the 
instantaneous luminosity recently reached the $10^{32}$cm$^{-2}$s$^{-1}$ 
level.
\end{itemize}
The frameworks in which these results will be presented are:
\begin{itemize}
\item most of the time ``standard SUSY''. This means the MSSM, often with some
generic unification constraints (essentially for slepton or gaugino 
SUSY-breaking masses) at the scale of grand unification (GUT), and with the 
assumption that the lightest 
supersymmetric particle (LSP) is the lightest neutralino $\chi$. Occasionnally,
the model considered will be even more constrained in the form of minimal 
supergravity (mSUGRA); 
\item in a few instances gauge mediated supersymmetry breaking (GMSB), which
offers clean and simple signatures, in its minimal version;
\item with R-parity conservation. Again, this is mostly due to lack of time 
in view of the large number of equally acceptable scenarios, and also partly 
because of a personal prejudice having to do with the absence of dark matter
in R-parity violating models. (Here I have to seek forgiveness from my HERA 
colleagues.)
\end{itemize}

At LEP, all supersymmetric particles, except for the gluinos, are produced in a
rather democratic fashion via electroweak interactions, up to mass effects. 
Since pair production of the LSP is not directly detectable, the search is 
naturally directed towards the next-to-lightest supersymmetric particles 
(NLSP's). The results can often be presented in a model independent, or at 
least moderately dependent, way. Results from various channels can furthermore 
be combined within some specific theoretical framework to derive additional 
constraints, of which the most celebrated one is the LSP-mass lower limit.

At the Tevatron, colored particles (i.e., squarks and gluinos) are expected to 
be produced with large cross sections via strong interactions. The resulting
final states consisting mostly of jets and missing transverse energy however 
suffer from large backgrounds from standard multijet production. The 
electroweak production of charginos and neutralinos has a much lower cross 
section, but clean final states such as trileptons and missing transverse 
energy are considerably easier to discriminate from standard model backgrounds.
Finally, in contrast to the situation at LEP, a model independent presentation 
of the results is usually unavailable because it is highly unpractical to 
provide one which would be both transparent and meaningful. 

All limits (unfortunately...) quoted in the following are given at 95\%
confidence level.  

\section{Standard SUSY: the LEP legacy}

\begin{figure}[htbp]
\begin{center}
\begin{tabular}{cc}
\raisebox{0.5cm}{
\mbox{\epsfig{file=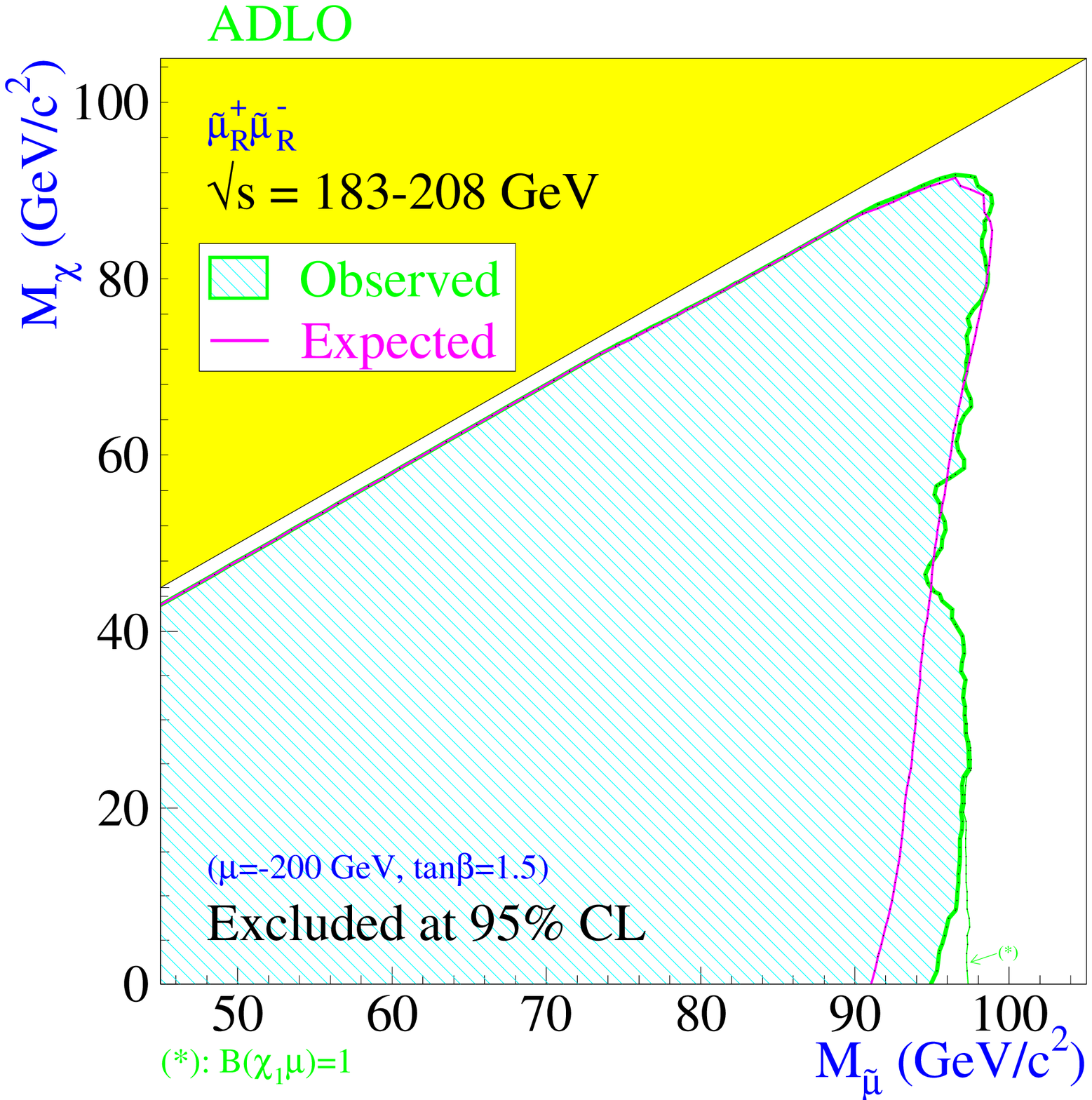,width=5.5cm}} }&
\raisebox{-0.5cm}{
\mbox{\epsfig{file=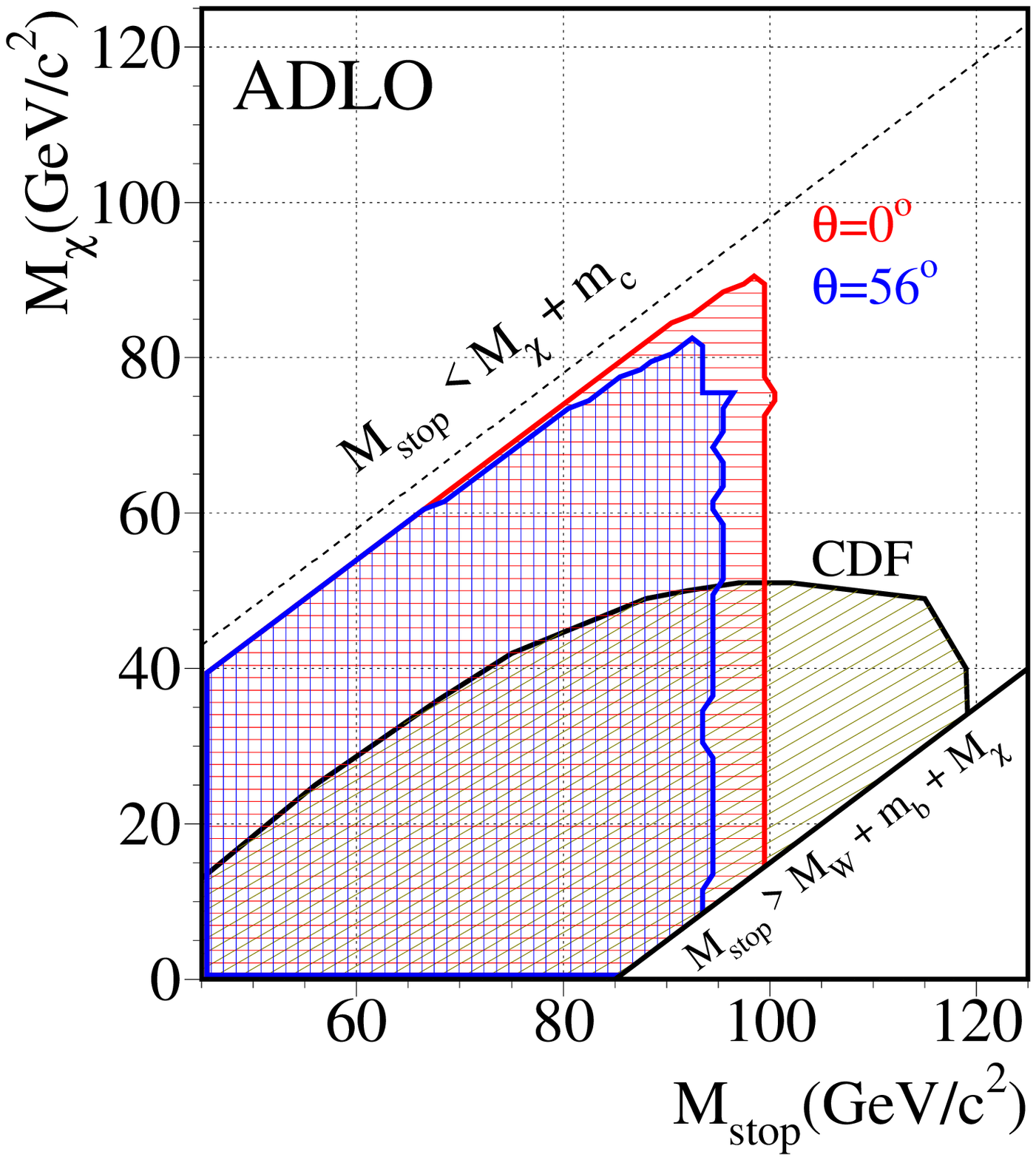,width=6.5cm}} }
\end{tabular}
\end{center}
\caption{\small{Domains excluded by the four LEP experiments: in the 
($m_{\tilde\mu},m_\chi$) plane (left), where the reduced sensitivity due to 
cascade decays is visible for low values of $m_\chi$; in the 
($m_{\tilde t},m_\chi$) plane (right), for $\tilde t\to c\chi$ prompt decays, 
where the innermost and outermost contours correspond to vanishing and 
maximum $Z\tilde t\tilde t$ couplings, and where the domain excluded by 
CDF at Run I is also shown.}}
\label{smuonstop}
\end{figure}

Because this is probably the simplest channel, both theoretically and 
experimentally, among those which have been analysed at LEP, let us begin 
with the search for smuon pair production, which proceeds only through 
$\gamma/Z$ exchange in the $s$-channel. It is assumed that it is the 
supersymmetric partner $\tilde\mu_R$ of the right-handed muon
which is the lighter of the two smuons, an assumption which is valid in models
involving the unification of SUSY-breaking masses and which is conservative
in terms of production cross section. The only parameter which is needed to 
calculate this cross section is the smuon mass. If the smuon is furthermore 
assumed to be the NLSP, the only decay channel available is 
$\tilde\mu_R\to\mu\chi$, and the only additional parameter involved is the 
LSP mass $m_\chi$. The final state consists in a pair of acoplanar muons, and 
the main background, from $WW\to\mu\nu\mu\nu$, is well under control. The
search result obtained by the four LEP experiments combined is shown in 
Fig.~\ref{smuonstop}(left)~\cite{LEPSUSY}; 
the gap along the diagonal is due to the 
softness of the muons when the $\tilde\mu$ -- $\chi$ mass difference is very 
small. With the additional condition of gaugino mass unification, the 
assumption that the smuon is the NLSP can be relaxed, and the effect of 
cascade decays such as $\tilde\mu\to\mu\chi'$ with $\chi'\to\gamma\chi$ can
be incorporated. These cascade decays occur only at small values of $m_\chi$, 
for which the smuon mass limit of almost 100~GeV is slightly degraded.

The interpretation of the other slepton searches involves some degree of model 
dependence. For staus, the mass eigenstates may significantly differ from the
electroweak eigenstates (in a way similar to stops, as discussed 
further down), and the most conservative results are quoted for the
case where the coupling of the lighter stau to the $Z$ vanishes~\cite{LEPSUSY}.
For selectrons, the situation is complicated by the contribution of $t$-channel
neutralino exchange to the pair production process. The interference is however
typically constructive, and the mass limits for selectrons are therefore 
slightly higher than for smuons~\cite{LEPSUSY}.

The mass matrix of scalar leptons or quarks contains off-diagonal terms which 
are proportional to the standard lepton or quark mass. Because of the large 
top-quark mass, it can be expected that mixing is important in the stop sector,
and therefore that the lighter stop could be substantially lighter than generic
squarks (and similarly, the lighter stau could be lighter than the other 
sleptons). This effect may even be enhanced in models with squark mass 
unification by the effect of the negative contribution of the large top-quark 
Yukawa coupling in the renormalization group equations for the stop masses. 
Although generic squarks are more efficiently searched at the Tevatron, well
beyond the kinematic reach of LEP, the lighter stop might still be light 
enough to be accessible at LEP. Given the chargino mass limits (discussed
further down), this stop would decay according to $\tilde t\to c\chi$. Since
this is a loop decay, the hadronization time may be similar to or shorter than
the decay time, and dedicated generators had to be set up to simulate the 
creation of stop-hadrons and their interaction with the detector. All
possible topologies arising from stop pair production have been analysed, 
ranging from acoplanar jets for short stop-hadron lifetimes, to pairs of heavy
charged or neutral particles for very long lifetimes, where the stop-hadron 
lifetime depends essentially on the $\tilde t$ -- $\chi$ mass 
difference~\cite{ALEPHstop}. 
For stop masses less than about 100~GeV, the domain excluded at the 
Tevatron is extended for $\tilde t$ -- $\chi$ mass differences smaller than 
$\sim$ 30~GeV, as shown in Fig.~\ref{smuonstop}(right)~\cite{LEPSUSY}.

Charginos can be pair produced by $s$-channel $\gamma/Z$ exchange, and by 
$t$-channel exchange of the electronic sneutrino, with a destructive
interference between the two channels. Similarly, pair or associated production
of neutralinos involves $Z$ and selectron exchanges, but now with a 
constructive interference. The analysis is greatly simplified under the
assumption that all sleptons are very heavy. In such a case, the production 
cross section for charginos depends on the three parameters entering the
chargino mass matrix: the SUSY-breaking gaugino mass $M_2$, the higgsino
mass term $\mu$, and $\tan\beta$, the ratio of the vacuum expectation values 
of the two Higgs fields. For neutralinos, an additional gaugino mass term is 
needed, $M_1$, with a value most commonly dictated by the unification 
condition ($M_1\sim M_2/2$).
 
For heavy sleptons, the decay processes are $\chi^+\to\chi W^\ast$ 
and $\chi'\to\chi Z^\ast$, so that the final states simply reflect the $W$
and $Z$ decay modes. All relevant topologies have been analysed: multijets,
acoplanar jets or leptons, leptons $+$ jets, all with missing energy carried 
away by the two LSP's. In the end, the kinematic limit of almost 104~GeV is 
essentially reached for chargino pair production~\cite{LEPSUSY}. 
For any fixed $\tan\beta$, 
this result can be turned into an exclusion domain in the ($M_2$,$\mu$) plane. 
This domain is slightly extended by the searches for neutralinos, which 
translates in turn into chargino exclusions indirectly extended by up to 
10~GeV (for low negative values of $\mu$ and small $\tan\beta$). The above
mentioned chargino searches loose sensitivity for very large values of $M_2$, 
because ot the correspondingly small $\chi^+$ -- $\chi$ mass difference. This
efficiency loss is however partially recovered using techniques such as the 
tagging of the production of an almost invisible final state by a photon from 
initial state radiation, or such as the identification of heavy stable charged 
particles when the $\chi^+$ -- $\chi$ mass difference becomes similar to or
smaller than the pion mass~\cite{LEPSUSY}. 
This last configuration is also encountered in 
models based on anomaly mediated supersymmetry breaking (AMSB).

Light sleptons have a negative impact on the sensitivity of the above searches.
The chargino production cross section is reduced, and this is not fully 
compensated by an increase of the neutralino production. Furthermore, the decay
branching ratios are modified, with a larger contribution of the leptonic final
states. In particular, invisible final states open up such as 
$\chi'\to\tilde\nu\nu$, or quasi-invisible ones such as 
$\chi^+\to\ell\tilde\nu$ when the $\chi^+$ -- $\tilde\nu$ mass difference is 
very small; the latter configuration has been dubbed ``the corridor''. The 
rescue comes from the direct searches for sleptons. With the assumption of a 
unified slepton mass $m_0$ at the GUT scale, a slepton mass limit can be 
translated into constraints in the ($m_0$,$M_2$) plane, for fixed $\tan\beta$. 
These constraints restrict in turn the allowed sneutrino masses, 
thus alleviating the negative impact of the corridor. 

\begin{figure}[htbp]
\begin{center}
\begin{tabular}{c}
\mbox{\epsfig{file=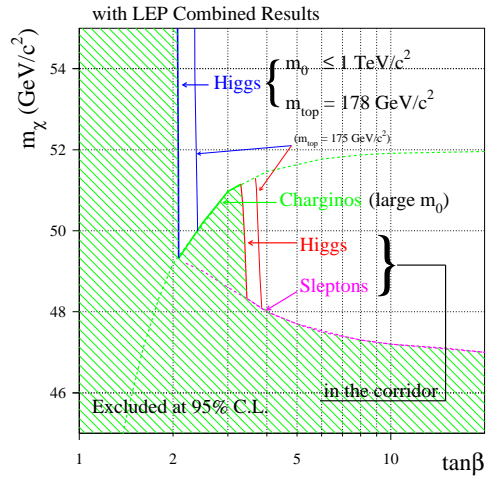,width=7cm}} 
\end{tabular}
\end{center}
\caption{\small{Limit on the mass of the lightest neutralino, as a function 
of $\tan\beta$, obtained by the four LEP experiments. From low to high values 
of $\tan\beta$, the exclusion is given by Higgs boson searches, by chargino 
searches, by Higgs boson searches in the corridor, by slepton searches in the
corridor.}}
\label{LSP}
\end{figure}

The power of the combination of various search channels is even more apparent
in the context of the limit which can be set on the mass of the LSP. It should 
however be emphasized that the results which will be given below are valid only
under the assumptions of slepton and gaugino mass unification at the GUT scale.
(Indeed, without gaugino mass unification, the neutralino-LSP mass could not be
constrained at all by LEP results.) The results are summarized in 
Fig.~\ref{LSP}~\cite{LEPSUSY}. For large slepton masses, the
limit on $m_\chi$ is about half the chargino mass limit for large $\tan\beta$,
somewhat lower otherwise. For low slepton masses, the lowest limit is set at
large $\tan\beta$ by selectron searches in the corridor. Finally, Higgs search
results are used to constrain the low $\tan\beta$ regime. In the end, the 
absolute LSP mass limit is 47~GeV.  

The constraints from Higgs boson searches were obtained as follows: 
sfermion (slepton and squark) and gaugino mass unifications are assumed; 
for a choice of $m_0$, $M_2$ and $\tan\beta$, the masses of $\tilde t_L$ 
and $\tilde t_R$ are computed, using the renormalization group equations; 
the mixing in the stop sector is set so as to maximize its impact on the 
Higgs boson mass; the mass $m_A$ of the pseudoscalar Higgs boson is 
chosen very large for the same reason; in this configuration, the
Higgs boson is standard model-like; if its expected mass $m_h$ is smaller 
than the combined LEP limit, the set of values considered for $m_0$, $M_2$ and 
$\tan\beta$ is excluded; a lower limit on $M_2$, hence on $m_\chi$, is 
obtained from a scan on $M_2$ (This limit is tightest for small values of 
$m_0$ and $\tan\beta$.); finally, a scan on $m_0$ allows an absolute 
$\chi$-mass lower limit to be set for the chosen value of $\tan\beta$.  
This method relies on the assumption that all $m_h$ values smaller than the 
LEP limit are excluded. There are however parameter configurations leading to
a Higgs boson mass smaller than the LEP limit, and for which the standard 
model Higgs boson searches are not sensitive (e.g., because of a vanishing 
branching ratio for $h\to b\bar b$). It has been verified by a fine scan of the
parameter space that all these configurations are indeed excluded by searches 
for supersymmetric particles, for charged or invisible Higgs bosons, for 
associated $hA$ production, or by flavor independent Higgs boson searches,
including dichotomies when appropriate (e.g., if two consecutive parameter 
sets are excluded by different searches)~\cite{ALEPHLSP}. 
In the end, the LEP limit was found
to be robust, and hence was appropriately used to set the LSP mass limit. The
result was also checked to be unaffected by variations of the top quark mass 
within its error.
 
The LSP mass limit was additionnally derived within the more constrained mSUGRA
framework. It was found to be slightly tighter, at the level of 
50~GeV~\cite{LEPSUSY}.

\section{A bit of Gauge Mediated SUSY Breaking}

In models based on GMSB, the SUSY breaking scale is much lower than in models
based on supergravity, so that the LSP is a very light gravitino $\tilde G$. 
The phenomenology depends largely on the nature and on the lifetime of the 
next-to-lightest SUSY particle (NLSP), which is a slepton (preferentially a 
stau) or the lightest neutralino $\chi$.

Pair production of stau-NLSP's has been searched by the four LEP experiments. 
Each stau decays to a $\tau$ and a gravitino, which can be considered massless 
for practical purposes. Depending on the stau lifetime, the final state evolves
from a pair of acoplanar taus, for which the standard SUSY search applies, to
a pair of stable massive charged particles. For intermediate lifetimes, 
dedicated searches for tracks with a large impact parameter or exhibiting a 
distinct kink were performed. In the end, stau-NLSP's lighter than 87~GeV are
excluded, independent of the stau lifetime~\cite{LEPSUSY}.

\begin{figure}[htbp]
\begin{center}
\begin{tabular}{c}
\mbox{\epsfig{file=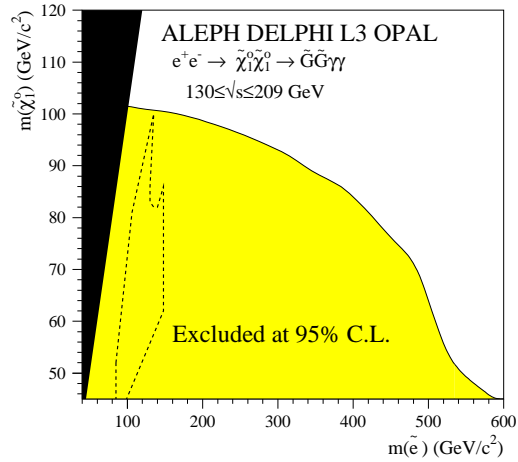,width=7cm}} 
\end{tabular}
\end{center}
\caption{\small{Domain excluded by the four LEP experiments in the plane 
($m_{\tilde e},m_\chi$), in GMSB with a neutralino NLSP 
(prompt $\chi\to\tilde G\gamma$ decays). The region compatible 
with the kinematicics of the ``CDF event'' is seen to be excluded.}}
\label{LEPGMSB}
\end{figure}

The pair production of $\chi$-NLSP's leads at LEP to a final state consisting
of an acoplanar photon pair ($\chi\to\tilde G\gamma$). No such signal was 
observed, leading to the exclusion of the domain shown in yellow in 
Fig.~\ref{LEPGMSB}~\cite{LEPSUSY}. 
For light selectrons, $\chi$ masses as high as 100~GeV are
excluded, which rules out the GMSB interpretation of a suggestive event 
observed by CDF in the Run I of the Tevatron, and containing two energetic 
electrons, two high energy photons, and a large missing transverse energy. 

\begin{figure}[htbp]
\begin{center}
\begin{tabular}{cc}
\raisebox{0.5cm}{
\mbox{\epsfig{file=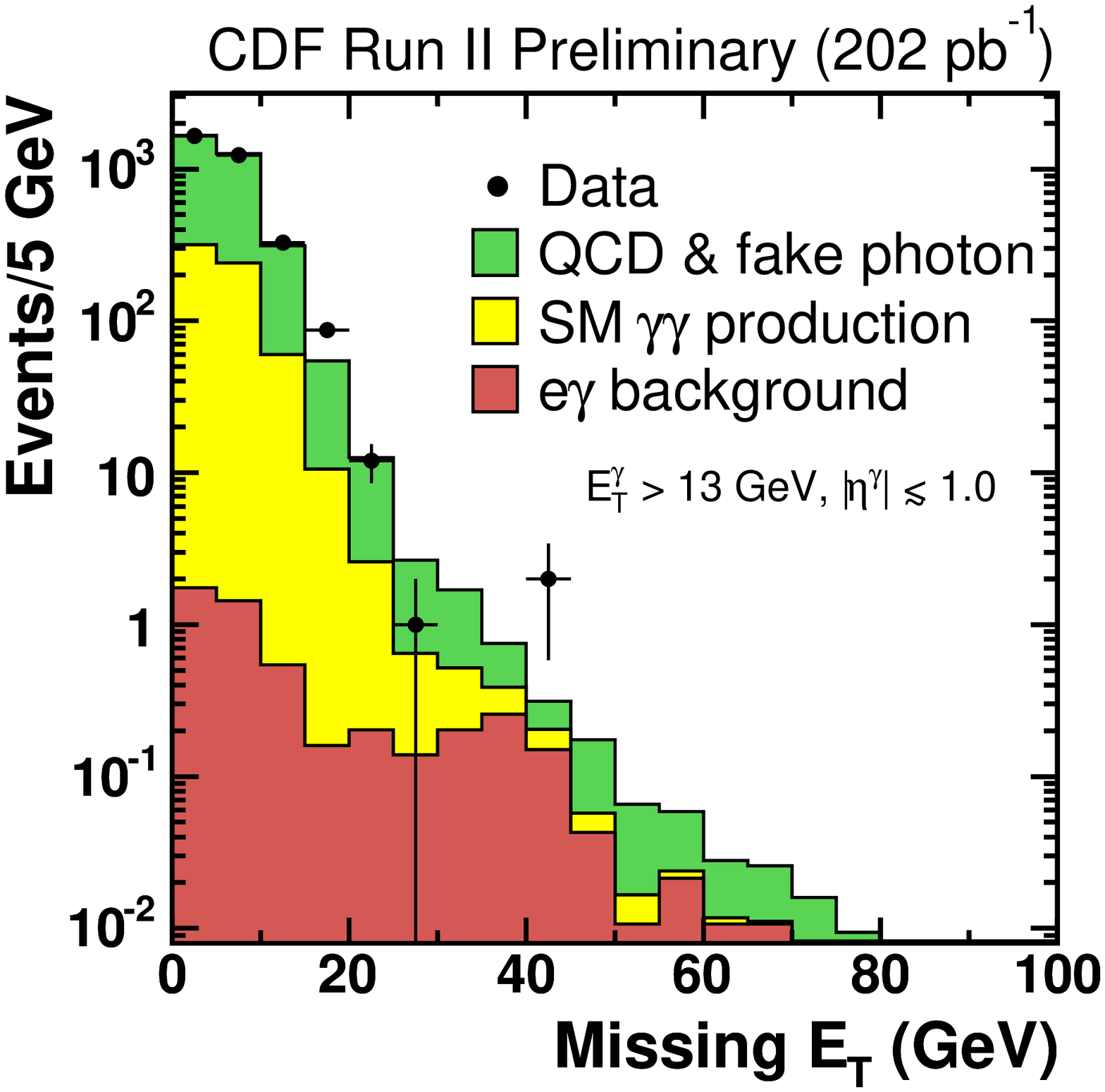,width=5.8cm}} } &
\raisebox{-0.5cm}{ 
\mbox{\epsfig{file=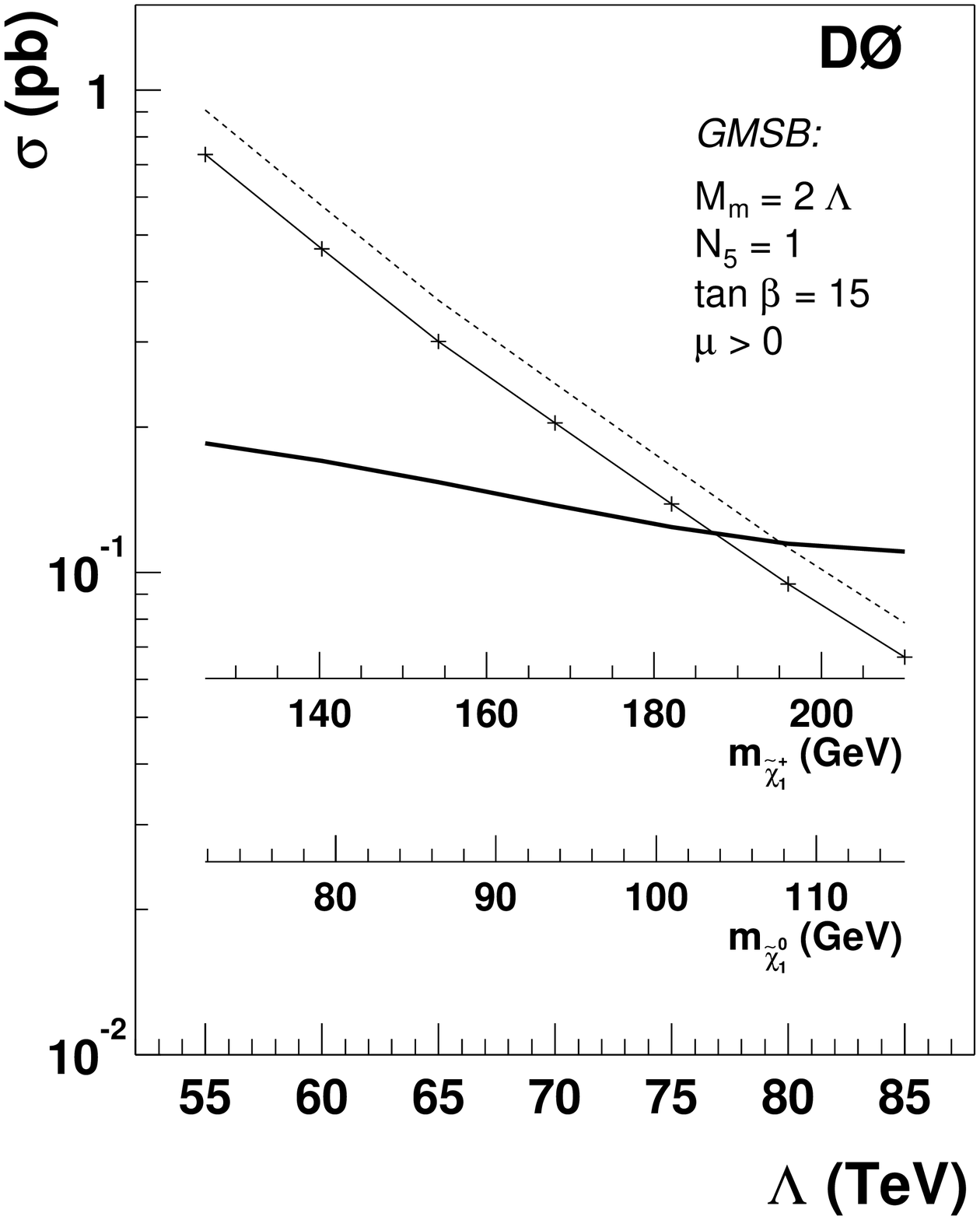,width=5.8cm}} }
\end{tabular}
\end{center}
\caption{\small{Tevatron searches for GMSB with a neutralino NLSP: the missing
$E_T$ distribution obtained by the CDF experiment in events containing two
high $E_T$ photons, with the various background contributions indicated (left);
the cross-section upper limit set by the D\O\ experiment, as a function
of the effective SUSY-breaking scale $\Lambda$ and of the $\chi$ mass, compared
to a theoretical expectation (thin dashed line) as detailed in the text 
(right).}}
\label{TEVGMSB}
\end{figure}

Searches for GMSB with a $\chi$-NLSP have been performed by both CDF and D\O\ 
in approximately 200 and 260~pb$^{-1}$ of Run II Tevatron data, respectively. 
These are inclusive
searches for topologies involving two photons and large missing transverse
energy $\met$. After mild quality and topological cuts, the main backgrounds
remaining are due to
\begin{itemize}
\item jets with a large electromagnetic component, thus faking photons, and to 
the QCD production of a photon and such a jet (or of two photons). Here the
missing $E_T$ is due to mismeasurements;
\item events containing real $\met$, with an electron falsely identified as a
photon, e.g. because the associated charged track was not reconstructed.
\end{itemize}
Both types of background are determined from the data. For the first one, 
events with reversed photon quality cuts are used to determine the shape of 
the $\met$ distribution, which is then normalized to the diphoton sample at 
small $\met$ values. The background with real electrons is evaluated requiring
that one of the would be photons is actually matched to a reconstructed track,
and folding in the track reconstruction and matching inefficiencies. The 
photon $p_T$ cuts are set to 13 and 20~GeV in CDF and D\O, respectively. The
resulting missing $E_T$ distribution
obtained by CDF is shown in Fig.~\ref{TEVGMSB}(left), together
with the various background contributions~\cite{CDF}. 
Optimal cuts on $\met$ are next 
applied, 45~GeV in CDF and 40~GeV in D\O, leading to 0 and 2 events observed,
respectively; the corresponding standard model expectations are 0.6 and 3.7 
events.

The interpretation of these results is performed within the minimal version of
GMSB, with the following choice of parameters: one set of messengers; the
messenger mass equal to twice the effective SUSY-breaking scale $\Lambda$; 
$\mu$ positive; and $\tan\beta = 15$. 
(This set corresponds to the so-called ``Snowmass slope''). For such parameter 
choices, the dominant mechanisms are chargino pair and associated 
chargino-second neutralino productions. The comparison of the cross section 
excluded by D\O\ with the predictions of
this model is shown in Fig.~\ref{TEVGMSB}(right)~\cite{DZERO}. It can be 
seen that $\Lambda$ values smaller than $\sim 80$~TeV are excluded, 
which corresponds to a lower limit of 108 GeV for the mass of the lightest 
neutralino, which is currently the world's most constraining result. 

\section{SUSY trileptons at the Tevatron}

The cleanest signature for SUSY at the Tevatron is expected to arise from
associated chargino-second
neutralino production, $p\bar p\to\chi^+\chi'$, followed
by leptonic decays, $\chi^+\to\chi\ell^+\nu$ and $\chi'\to\chi\ell^+\ell^-$,
in which case the final state contains three leptons and missing $E_T$. The
drawbacks are that the cross sections (times branching ratios) are rather 
small, that the final state leptons have soft spectra, and that final states
containing $\tau$'s can be enhanced, in particular at large $\tan\beta$. 
A large integrated luminosity is therefore needed, and various trilepton final
states have to be combined to increase the search sensitivity.

The D\O\ collaboration has investigated trilepton toplogies in 145 to 
250~pb$^{-1}$ of Run II data, depending on the final state analysed. The
signatures considered up to now were $ee\ell$, $e\mu\ell$, $\mu\mu\ell$
and same sign 
dimuons, where the third lepton $\ell$ is not required to be positively 
identified, which increases the efficiency for final states involving $\tau$'s.
The analysis was optimized for chargino and neutralino masses just beyond the 
LEP limits, and for slepton masses just above the $\chi'$ mass in order to
enhance leptonic branching ratios. The mSUGRA framework (with $\tan\beta=3$, 
$A_0=0$ and $\mu>0$) was used to make definite predictions, but the results 
are expected to hold more generally in the MSSM when the mass hierarchy 
$m_{\chi^+} \sim m_{\chi'} \sim 2m_{\chi} \sim m_{\tilde\ell}$ is preserved.

\begin{figure}[htbp]
\begin{center}
\begin{tabular}{c}
\mbox{\epsfig{file=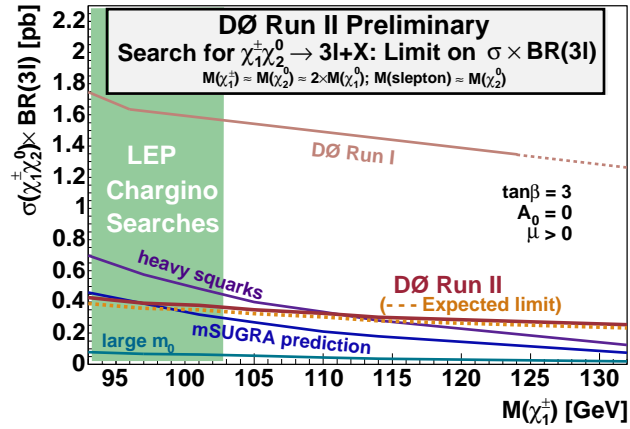,width=9cm}} 
\end{tabular}
\end{center}
\caption{\small{Upper limit on the product of cross section and branching ratio
into three leptons obtained by the D\O\ experiment, as a function of the 
chargino mass. The results are compared to an mSUGRA theoretical prediction. 
The region excluded by the LEP chargino searches is also shown.}}
\label{trilep}
\end{figure}

The selections require two identified, possibly rather soft, isolated leptons 
(electrons or muons), some significant missing $E_T$, and reject events where 
a pair of opposite sign electrons or muons is compatible with originating from 
a $Z$-boson decay.
Backgrounds are further reduced by the requirements of either a same sign for
two muons, or of the presence of an additional isolated charged particle track
(the ``third'' lepton). Backgrounds originating from $WW$, $WZ$ and $W\gamma$ 
(where the photon converts into an $e^+e^-$ pair in the detector) are estimated
by simulation, while the background from $b\bar b$ production, which may lead
to moderately isolated muons, is determined using data. In the end, three
candidate events are selected, while the expected background amounts to 
$2.9 \pm 0.8$ events. This result translates into an upper limit for the
product of the production cross section by the leptonic branching ratio, 
as shown in Fig.~\ref{trilep}~\cite{DZERO}. 
The improvement over the D\O\ Run I result is 
quite substantial, but not yet sufficient to reach the level of the mSUGRA
prediction. An increased integrated luminosity and the inclusion of a search
for same sign electron pairs should
allow mSUGRA territory to be entered in the near future. 

\section{Squarks and gluinos at the Tevatron}

The CDF collaboration has performed a search for stable massive charged 
particles in 53~pb$^{-1}$ of Run~II data. Such particles are expected to 
behave like slow moving heavy muons in the detector, and can be identified 
using a time of flight technique. The null result, interpreted in terms of
absence of stable stop production, leads to mass lower limits of 108~GeV if
such stable stops appear isolated, and of 95~GeV otherwise~\cite{CDF}.

\begin{figure}[htbp]
\begin{center}
\begin{tabular}{c}
\mbox{\epsfig{file=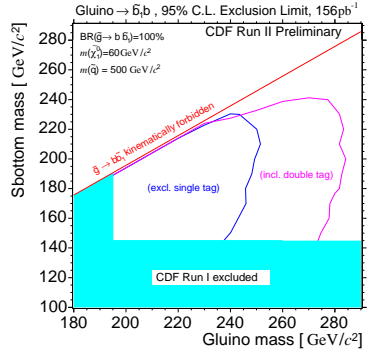,width=5cm}} 
\end{tabular}
\end{center}
\caption{\small{Domain excluded by the CDF collaboration in the plane 
($m_{\tilde g},m_{\tilde b}$), for light gluinos and sbottom quarks, and 
for $m_\chi = 60$~GeV.}}
\label{CDFsbot}
\end{figure}

For large values of $\tan\beta$, mixing is large in the sbottom sector, and
the mass hierarchy could well be such that gluinos are abundantly produced, 
and always decay into a $b\tilde b$ pair. A light sbottom is expected to decay
according to $\tilde b\to b\chi$, so that the final state resulting from 
gluino pair production would contain four $b$ quarks and exhibit missing 
$E_T$. The CDF collaboration has searched for this topology in 156~pb$^{-1}$
of Run~II data, requiring one or two of the four jets to be $b$-tagged. 
The second option gives the best sensitivity, and typically allows gluinos 
with masses up to 280~GeV to be excluded for sbottom masses smaller than 
240~GeV (and for $m_\chi = 60$~GeV) , within the specific theoretical 
framework considered (Fig.~\ref{CDFsbot})~\cite{CDF}.

Generic squark production has been investigated by the D\O\ collaboration,
using 85~pb$^{-1}$ of Run~II data. If squarks are lighter than gluinos, they
are expected to decay according to $\tilde q\to q\chi$, in which case the 
final state arising from squark pair production is a pair of acoplanar jets.
Other squark ``cascade'' decay modes may however spoil this topology, such as 
$\tilde q\to q\chi'$ or $q'\chi^{\pm}$. Since the decay branching ratios are 
model dependent, the mSUGRA framework was used to conduct the analysis, with 
$\tan\beta=3$, $A_0=0$, and $\mu<0$. The model line $m_0 = 25$~GeV was chosen
so as to lead to squarks essentially as light as possible, and the only free
parameter is therefore $M_2$, which controls simultaneously the squark and 
gluino masses.

\begin{figure}[htbp]
\begin{center}
\begin{tabular}{cc}
\mbox{\epsfig{file=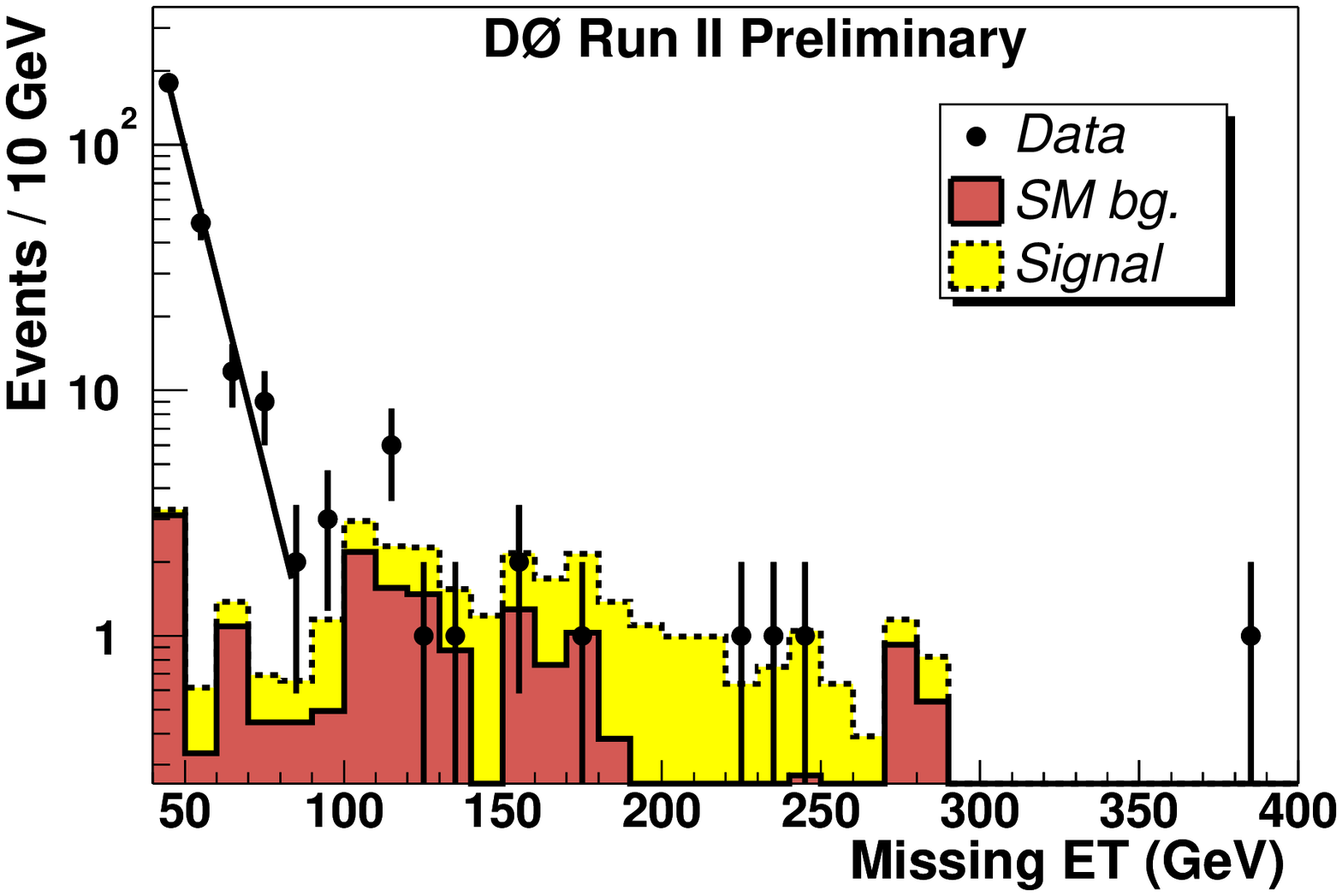,width=7cm}} & 
\raisebox{.3cm}{
\mbox{\epsfig{file=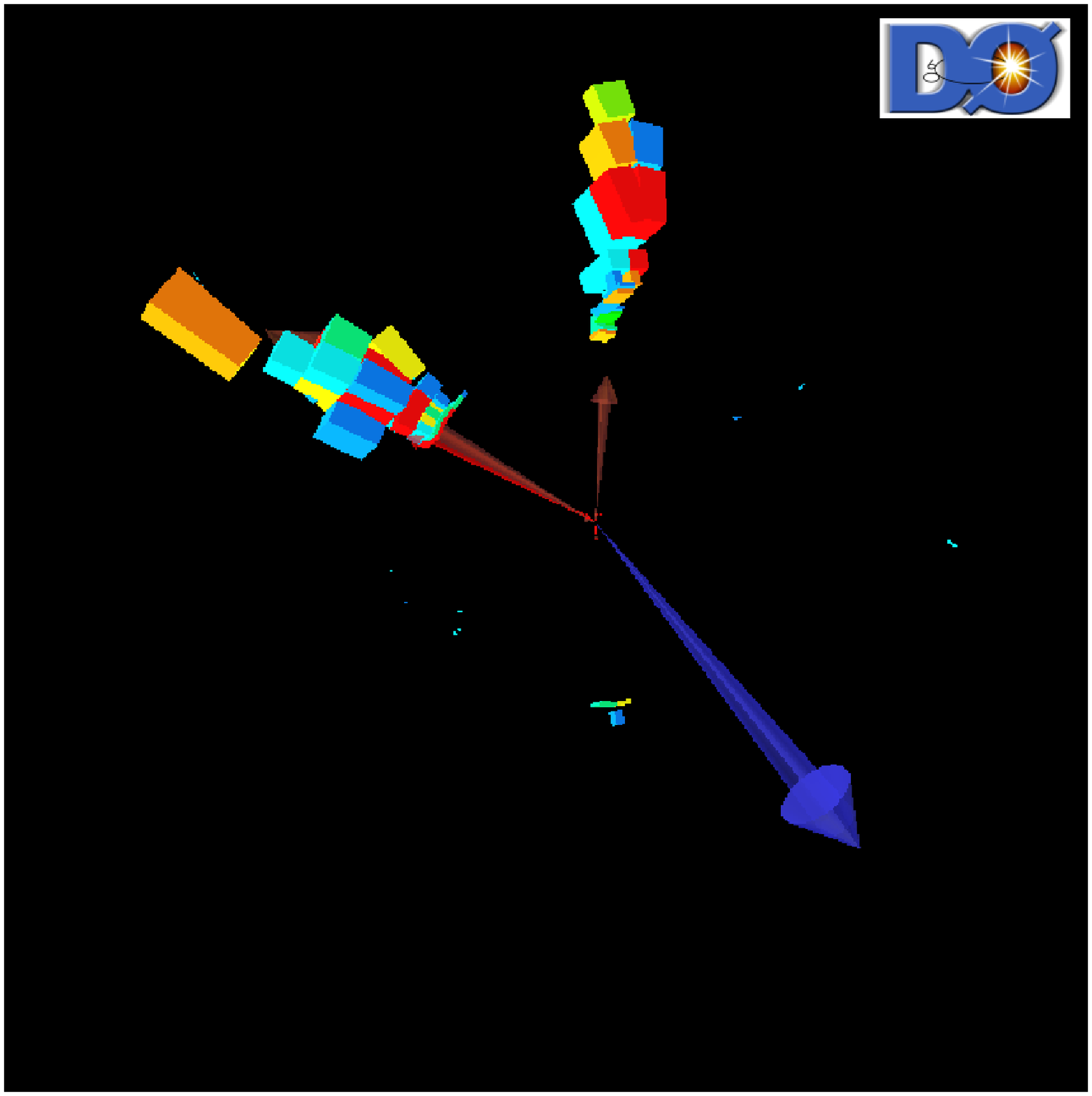,width=4.5cm}} }
\end{tabular}
\end{center}
\caption{\small{Search for generic squarks by the D\O\ collaboration: the 
final missing $E_T$ distribution (left), with the contributions from standard
model processes (brown), from QCD (exponential fit), and expected from the
signal (yellow, on top of standard model); 
three-dimensional view of the highest missing $E_T$ event (right).}}
\label{squarks}
\end{figure}

The main selection cuts were designed to select pairs of acoplanar jets while 
rejecting as much as possible of the main backgrounds: multijet production from
standard QCD processes, with jet energy mismeasurements creating fake $\met$; 
and associated production of a $W$ boson and jets, with $W\to\ell\nu$. The 
associated production of a $Z$ boson and two jets causes an irreducible 
background when the $Z$ decays into a $\nu\bar\nu$ pair. Two high $p_T$ jets 
were required and a veto against isolated leptons was applied. The missing 
$E_T$ was required not to be directed along (or opposite to) any jet. The final
cuts at 275~GeV on the sum $H_T$ of all jet transverse energies and at 175~GeV
on the missing $E_T$ were optimized to give the smallest signal production 
cross section expected to be excluded at the edge of the Run~I exclusion 
domain. The missing $E_T$ distribution is shown in Fig.~\ref{squarks}(left). 
It can be
seen that the QCD background, clearly visible at low $\met$, is negligible
beyond the chosen $\met$ cut value. Four events were selected in the data, 
while
$(2.7 ^{+2.3}_{-1.5})$ are expected from standard model processes, mostly from
$Z\to\nu\bar\nu$ and from $W\to\tau\nu$. The highest $\met$ event is shown in
Fig.~\ref{squarks}(right).
Along the model line chosen, squark masses smaller than 292~GeV are excluded
(assuming four mass-degenerate squark species), as well as gluino masses up to
333~GeV~\cite{DZERO}. 
These results slightly improve on those obtained at Run~I.

The question can be raised of the relevance of the Tevatron searches for 
squarks and gluinos. Indeed, if gaugino mass unification is assumed, as it is 
the case in mSUGRA, the LEP limits on charginos translate into gluino mass 
limits well beyond those within the Tevatron reach. Similarly, if slepton and
squark mass unification is assumed, the LEP limits on selectrons are more
restrictive than those on squark masses from the Tevatron. The CDF and D\O\
collaborations should therefore be encouraged to present their results within
frameworks other than mSUGRA. For instance, in SUSY-GUT models where SUSY
breaking is induced by an F-term which is a {\bf 75} of SU(5), rather than by 
a singlet, the $M_3/M_2$ ratio is of order unity, in which case LEP results do 
not constrain gluino masses above 105~GeV or so. String inspired models can 
also lead to similar gaugino mass hierarchies.   

\section{Conclusions}

Before concluding, I want to quote a result which is a bit off the main track,
but still relevant for supersymmetry. The CDF and D\O\ collaborations have 
searched for the rare decay $B_s\to\mu^+\mu^-$ which is expected to be at the
3.5~10$^{-9}$ level in the standard model, but may be enhanced by a factor of
$(\tan\beta)^6$, i.e., by as much as three orders of magnitude, 
in supersymmetry.
At the time this talk was given, the CDF collaboration had quoted a limit of
7.5~10$^{-7}$~\cite{CDF}, 
improving substantially over the previous best limit and probing
relevant new territory, while the D\O\ collaboration had not yet ``opened the 
box''.\footnote{Since then, the box has been opened, and a limit of 
5.0~10$^{-7}$ has been set~\cite{DZB}} 

As of today, the main constraints on supersymmetry obtained at accelerators 
remain those established by LEP. If fine tuned parameter configurations are 
discarded, limits of the order of 100~GeV are set on slepton and 
chargino masses, and the mass of the lightest supersymmetric particle has to 
exceed 47~GeV in the MSSM with slepton, squark and gaugino mass unification.
 
The Tevatron is however already providing relevant results. In the framework 
of gauge mediated supersymmetry, a lower limit of 108 GeV has been set on the 
mass of a  neutralino NLSP. Trilepton searches should lead to new constraints 
on minimal supergravity in the near future. Squark and gluino searches are well
underway, although an adequate interpretation of the results is still lacking.
With the continuously improving performance of the Tevatron, the coming years 
can be expected to provide an exciting harvest of new results.

\end{document}